\documentclass[aps,prb,twocolumn,superscriptaddress,showpacs]{revtex4}
\usepackage{graphicx}
\usepackage{calc}
\usepackage{bm}

\begin{document}

\title{Charge redistribution between cyclotron-resolved edge states at high imbalance}

\author{E.V.~Deviatov}
\email[Corresponding author. E-mail:~]{dev@issp.ac.ru}
\affiliation{Institute of Solid State Physics RAS, Chernogolovka,
Moscow District 142432, Russia}

\author{V.T.~Dolgopolov}
\affiliation{Institute of Solid State Physics RAS, Chernogolovka,
Moscow District 142432, Russia}

\author{A.~W\"urtz}
\affiliation{Laboratorium f\"ur Festk\"orperphysik, Universit\"at
Duisburg-Essen, Lotharstr. 1, D-47048 Duisburg, Germany}

\date{\today}

\begin{abstract}
We use a quasi-Corbino sample geometry with independent contacts
to different edge states in the quantum Hall effect regime to
investigate a charge redistribution between cyclotron-split edge
states at high imbalance. We also modify  B\"uttiker formalism by
introducing local transport characteristics in it and use this
modified B\"uttiker picture  to describe the experimental results.
We find that charge transfer  between cyclotron-split edge states
at high imbalance can be described by a single parameter, which is
a transferred between edge states portion of the available for
transfer part of the electrochemical potential imbalance. This
parameter is found to be independent of the particular sample
characteristics, describing fundamental properties of the
inter-edge-state scattering. From the experiment we obtain it in
the dependence on the voltage imbalance between edge states and
propose a qualitative explanation to the experimental findings.
\end{abstract}


\maketitle

Just from the beginning of the quantum Hall investigations it was
understood that edge states play a significant role in many
transport phenomena in the quantum Hall effect
regime~\cite{halperin}. In a quantizing magnetic field  the edge
potential bends up the energy levels near the sample edges. At the
intersections of the energy levels with  Fermi level edge states
are formed. It was a paper of B\"uttiker~\cite{buttiker} that
proposes a formalism for the Hall resistance calculation regarding
a transport through edge states. This model was further developed
by Chklovskii {\it et al.}~\cite{shklovsky} for electrostatically
interacting electrons. The interaction modifies one-dimentional
B\"uttiker edge states into the stripes of incompressible electron
liquid of finite widths. It was shown
theoretically~\cite{buttiker} and confirmed in
experiments~\cite{haug} that quantum Hall resistance is not
sensitive to the inter-edge-channel scattering. Nevertheless, the
properties of this scattering can be investigated by using the
selective edge channel population methods.

Most of experiments have been performed in the Hall-bar geometry
by using the cross gate technique~\cite{haug}. These experiments
have revealed the inter-edge-scattering dependence on the magnetic
field, temperature and filling factor~\cite{haug}. In the Hall-bar
geometry the experiments are at low imbalance conditions, when the
energy difference between edge states is smaller than the spectral
gaps. An attempt to increase the edge states imbalance by closing
cross-gates  dramatically decreases the experimental accuracy,  as
it was mentioned in Ref.~\onlinecite{muller}.

Another experimental method is the using of the quasi-Corbino
sample geometry~\cite{weiss,alida}. In this geometry two
not-connecting etched edges are formed in the sample. A cross-gate
is used to redirect some edge states between etched edges and to
define an interaction region at one edge. Because the interacting
edge states originate from different edges of the sample, they are
independently contacted and direct inter-edge-scattering
investigations become possible at any imbalance between edge
states. This imbalance is controlled by the applied voltage, and
in dependence of it's sign the edge potential profile between edge
states becomes stronger of flatter. In the later case at some
voltage imbalance the potential barrier between edge states
disappears, leading to a step-like behavior of the corresponding
branch of the $I-V$ curve. This effect opens a path to use the
quasi-Corbino geometry for spectroscopical investigations at the
sample edge. Recently, the quasi-Corbino geometry was used to
investigate the edge spectrum of single-\cite{alida} and
double-\cite{dqwedge} layer two-dimensional electron structures.
It was also understood that in the transport between spin-resolved
edge states at high imbalance (i.e. higher than the spectral gaps)
nuclear effects become important~\cite{DNP}.

\begin{figure}
\includegraphics*[width=0.5\columnwidth]{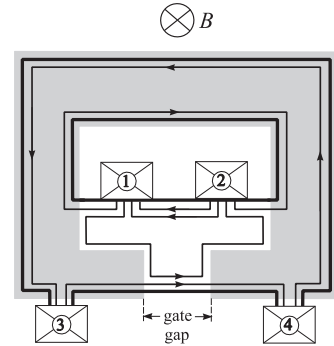}%
\caption{Schematic diagram of the pseudo-Corbino sample geometry.
Contacts are positioned along the etched edges of the ring-shaped
mesa (thick outline). The shaded area represents the
Schottky-gate. Arrows indicate the direction of electron drift in
the edge states. \label{sample}}
\end{figure}

When developed,  B\"uttiker formalism was intended to describe a
high accuracy of the sample resistance quantization in the quantum
Hall effect regime. For this reason, it depicts the inter-edge
scattering by integral sample characteristics, practically as
scattering between ohmic contacts. This picture becomes
inconvenient while describing a charge transfer between edge
states at high imbalance, where the scattering {\em by definition}
takes place on small lengths, much smaller than the sample size.

Here, we investigate  a charge transfer between cyclotron-split
edge states at high imbalance. We modify  B\"uttiker formalism by
introducing local transport characteristics in it. We find that
charge transfer can be described by a single parameter, which is
the transferred portion of the available for the transfer  part of
the electrochemical potential imbalance. This modified B\"uttiker
picture is used to describe details of charge transfer while
current is overflowing between edge channels.

Our samples are fabricated from a molecular beam
epitactically-grown GaAs/AlGaAs heterostructure. It contains a
two-dimensional electron gas (2DEG) located 70~nm below the
surface. The mobility at 4K is 800 000 cm$^{2}$/Vs and the carrier
density 3.7 $\cdot 10^{11}$cm$^{-2}$. Samples are patterned in a
quasi-Corbino geometry~\cite{alida}, see Fig.~\ref{sample}.
Rectangular mesa has an etched region inside.   Ohmic contacts are
made to both (inner and outer) edges of the sample. A Shottky gate
is patterned around the inner etched area, leaving uncovered
T-shaped region between inner and outer edges. This region forms a
narrow (about several microns) strip of uncovered 2DEG near the
outer edge of the sample which is called gate-gap.  Here we
present data from the sample with $5~\mu$m gate-gap width, while
$2,10,20~\mu$m gate-gap samples are also investigated showing
identical experimental results.

In our experimental set-up one of the inner contacts is always
grounded. In a quantizing magnetic field, at filling factors
$\nu=3,4$, we deplete 2DEG under the gate to a smaller filling
factor $g=2$, redirecting cyclotron-split $\nu-g$ edge states from
inner to outer edges of the sample. We apply a dc current to one
of the outer contacts and measure a dc voltage drop between two
others inner and outer contacts at the temperature of 30~mK.  By
switching current and voltage contacts $I-V$ traces for four
different contact combinations can be investigated. Because of
independent ohmic contacts to the cyclotron-split edge states, the
measured voltage $U$ is connected to the voltage drop $V$ between
edge states in the gate-gap, which is directly the energy shift
$eV$ between them. For example, $U=V$ for contact combination at
which contacts no. 4 and 2 are current contacts and no. 3 and 1
are voltage ones, as denoted in Fig.~\ref{sample}.

\begin{figure}
\includegraphics[width=\columnwidth]{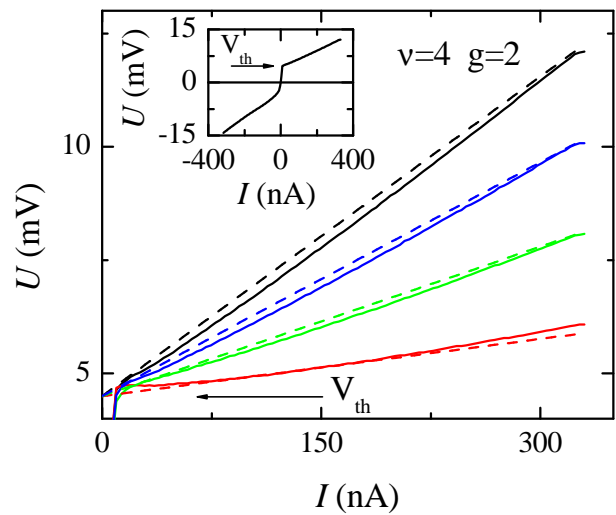}%
\caption{ Positive branches of experimental $I-V$ curves for
filling factors $\nu=4,g=2$ for different contact configurations.
They are (from up to down): current (4-2), voltage (3-1);
(4-1),(3-2); (3-2),(4-1); (3-1),(4-2) as depicted in
Fig.\protect~\ref{sample}. The inset shows an example of the
experimental $I-V$ curve in a whole sweeping range with marked
threshold position. $V_{th}=4.5$~mV. The magnetic field is
$B=3.9$~T. \label{IV42}}
\end{figure}

\begin{figure}
\includegraphics[width=\columnwidth]{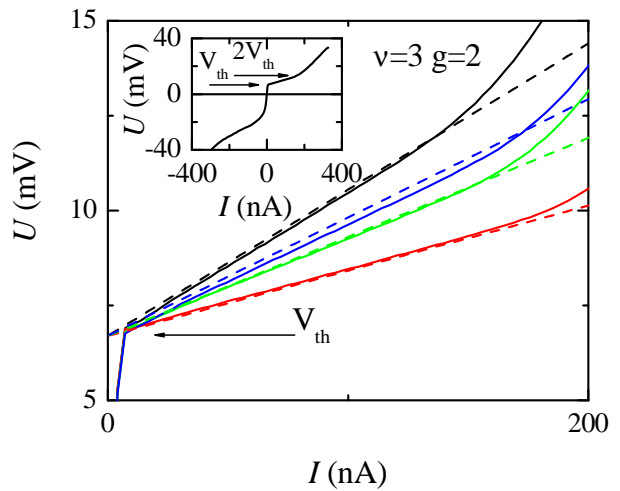}%
\caption{   Positive branches of experimental $I-V$ curves for
filling factors $\nu=3,g=2$ for different contact configurations.
They are (from up to down): current (4-2), voltage (3-1);
(4-1),(3-2); (3-2),(4-1); (3-1),(4-2) as depicted in
Fig.\protect~\ref{sample}. The inset shows an example of the
experimental $I-V$ curve in a whole sweeping range with marked
threshold and twice threshold positions. $V_{th}=6.7$~mV. The
magnetic field is $B=5.1$~T. \label{IV32}}
\end{figure}

Examples of experimental $I-V$ curves are presented in insets to
Figs.~\ref{IV42},\ref{IV32} for two groups of cyclotron-split edge
states. While increasing a current from zero to positive values,
the measured voltage rising abruptly to a some value $V_{th}$.  It
is practically no current before $V=V_{th}$, but after it the
voltage is a roughly linear function of the current. This linear
law is valid for hundreds of nanoAms, see main
Figs.~\ref{IV42},\ref{IV32}, up to our highest applied currents
for filling factor combination $\nu=4,g=2$. For $\nu=3,g=2$ at
high currents there is a strong deviation from the linear law. The
deviation starts from twice the onset voltage $2V_{th}$, and leads
to increasing the resistance in respect to the linear dependence.
It can not be due to  overheating the sample by the current
because it would diminish the resistance, in contradiction with
the experiment, see Fig.~\ref{IV32}.

In Figs.~\ref{IV42},\ref{IV32} positive $I-V$ branches are shown
for four different contact combinations. As it can be seen from
the figure, there is still small non-linearity of the curves. The
described above behavior is valid for all of them and is very
reproducible from sample to sample and  in cooling cycles.
Positive branches start from the same threshold voltage, which is
fixed for a given filling factor combination. The threshold
voltage values are close to the cyclotron splitting in the
corresponding field, but smaller on approximately 2~mV, see
Ref.~\onlinecite{alida}: $V_{th}=4.5$~mV for $\nu=4,g=2$ ($\hbar
\omega_c=6.7$~meV) and $V_{th}=6.7$~mV for $\nu=3,g=2$ ($\hbar
\omega_c=8.8$~meV).

While sweeping the current to the negative values, there is no a
clear defined onset: the voltage is rising with rising the current
practically from a zero value. The negative branch of the $I-V$
curve is clearly non-linear for any currents, see insets to
Figs.~\ref{IV42},\ref{IV32}. The exact form of the branch is
dependent on the cooling procedure and may variate from cycle to
cycle.

To be correct, B\"uttiker formalism~\cite{buttiker}  can not be
conveniently applied to the transport at high imbalance. It
describes {\em integral} sample resistance, so in the case of
non-linear $I-V$ curve the B\"uttiker transmission coefficients
become to be dependent on the voltage imbalance between edge
states.

As an example, let us consider a filling factor combination
$\nu=4,g=2$. Our sample can be described by the
equations~\cite{buttiker}:
\begin{eqnarray}
I_1=4\frac{e}{h}\mu_1-4\frac{e}{h}\mu_2, \nonumber\\
I_2=4\frac{e}{h}\mu_2-2\frac{e}{h}\mu_1-2\frac{e^2}{h}(T_{21}\mu_1+T_{23}\mu_3),\nonumber\\
I_3=2\frac{e}{h}\mu_3-2\frac{e}{h}\mu_4,\nonumber\\
I_4=2\frac{e}{h}\mu_4-2\frac{e}{h}(T_{41}\mu_1+T_{43}\mu_3),
\label{eqButSys}
\end{eqnarray}
where $I_i$ is the current flowing in the $i$-th contact, $\mu_i$
is the electrochemical potential of the $i$-th contact, and
$\{T_{ij}\}$ is the matrix of transmission
coefficients~\cite{buttiker}. This coefficients are not
independent: because of the charge conservation in the gate-gap we
can write
\begin{eqnarray}
T_{21}+T_{41}=1, \nonumber \\
T_{23}+T_{43}=1. \label{eqButT}
\end{eqnarray}
Also from  symmetry considerations we should mention that
$$T_{23}=T_{41}.$$ It means that every transmission coefficient
can be expressed through a single value, which we define as
$T=T_{23}$.

 Let the current  flow between contacts no. 4 and 1, and use the
contacts no. 3 and 2 to measure the voltage drop. For these
experimental conditions the flowing current is
$I_{41}=I_{1}=-I_{4}$ and there is no current in the voltage
probes $I_{2}=I_{3}=0$. Also the voltage drop is a difference of
the electrochemical potentials of the potential contacts,so
$eU_{32}=\mu_3-\mu_2$. By solving a system~(\ref{eqButSys}) with
relations~(\ref{eqButT}) and herein we can obtain
\begin{equation}
U_{32}=\frac{2-T}{4T}\frac{h}{e^2} I_{41}. \label{eqBut}
\end{equation}

\begin{figure}
\includegraphics[width=\columnwidth]{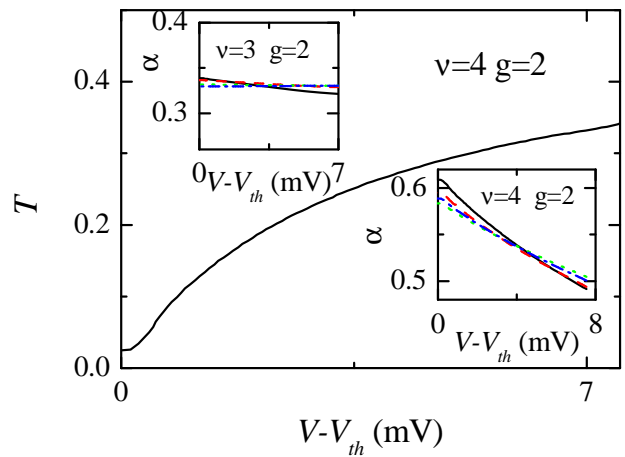}%
\caption{  The dependence of the B\"uttiker transmission
coefficient $T$ on the voltage imbalance between
cyclotron-resolved edge states at filling factors $\nu=4,g=2$,
starting from the threshold voltage. Insets show the dependencies
of the transport parameter $\alpha$ (see text) as obtained from
$I-V$ curves at four different contact configurations for
$\nu=3,g=2$ and $\nu=4,g=2$ filling factors correspondingly.
\label{But}}
\end{figure}

The relation~(\ref{eqBut}) can be used to calculate $T$ from the
experimental $I-V$ trace, see Fig.~\ref{But}. The dependence
$T(V)$ is strongly non-linear. It starts from the threshold
voltage, because below threshold practically no current is flowing
so the transmission $T$ is practically zero. While increasing the
voltage imbalance between edge states $V$, $T(V)$ is monotonically
rising and asymptotically trends to the equilibrium B\"uttiker
value $T=1/2$ at high voltages $V$. $T(V)$ dependance has a
universal character: while obtained, it can be used to describe
the experimental $I-V$ traces for any given contact combination at
fixed filling factors. One should calculate the current-voltage
relation for this contact combination from Eq.~(\ref{eqButSys})
and introduce the found above $T(V)$ into it to obtain the
experimental $I-V$ curve. We will demonstrate this fact below in a
physically more transparent manner.

Using strongly non-linear transparency $T(V)$ is too sophisticated
to analyze the overflowing current at $V>V_{th}$, e.g. it is not
clear the physical origin of the linear regions on the
experimental $I-V$ curves. For the non-linear transport in the
gate-gap it is obvious to introduce local transport
characteristics instead of the integral B\"uttiker transmission
coefficient $T$. From the positive branch of the experimental
$I-V$ curve we conclude that there is practically no current
between edge states below the threshold voltage. In B\"uttiker
picture of edge states it means that both edge states are
injecting and leaving the gate-gap region with their own
electrochemical potentials $\mu_1$ and $\mu_3$ , originating from
corresponding ohmic contacts no.~1 and no.~3. Currents, flowing in
the gate-gap are equal to $e/h \mu_1$ and $e/h \mu_3$ in the inner
and outer edge states correspondingly. A current {\em between}
edge states starts to flow then the difference in electrochemical
potentials exceeds the threshold voltage. In the other words, only
some part of the incoming electrochemical imbalance
$(\mu_3-\mu_1-eV_{th})$ is available for redistribution between
edge states. It is obvious in this case to describe the current
between edge states as $ \alpha (\mu_3-\mu_1-eV_{th})e/h$, where
$\alpha$ is a parameter, describing a portion of the available
part of electrochemical potential imbalance, which is {\em in
fact} transferred between edge states. For the described above
filling factor combination $\nu=4,g=2$ it is clear that
$\alpha=1/2$ means equal redistribution between edge states. The
edge states are leaving the gate-gap region with mixed
electrochemical potentials $\mu_1+\alpha (\mu_3-\mu_1-eV_{th})$
and $\mu_3-\alpha (\mu_3-\mu_1-eV_{th})$. By introducing these
values into the B\"uttiker formulas~(\ref{eqButSys}) we have the
following equations instead of (\ref{eqButSys}-\ref{eqBut}):
\begin{eqnarray}
I_1=4\frac{e}{h}\mu_1-4\frac{e}{h}\mu_2,\nonumber\\
I_2=4\frac{e}{h}\mu_2-2\frac{e}{h}\mu_1-2\frac{e}{h}(\mu_1+\alpha
(\mu_3-\mu_1-eV_{th})),\nonumber\\
I_3=2\frac{e}{h}\mu_3-2\frac{e}{h}\mu_4,\nonumber\\
I_4=2\frac{e}{h}\mu_4-2\frac{e}{h}(\mu_3-\alpha
(\mu_3-\mu_1-eV_{th})). \label{eqAlSys}
\end{eqnarray}
In this case there is no need in any additional relations (all the
necessary information is indeed in the equations~(\ref{eqAlSys}))
and the only parameter $\alpha$ has a clear physical sense: it is
a transferred between edge states portion of the available for the
transfer part of the electrochemical potential imbalance between
edge states. The above mentioned combination of filling factors
and contacts can be described by $I-V$ relation
\begin{equation}
U_{32}-V_{th}=\frac{2-\alpha}{4\alpha}\frac{h}{e^2} I_{41}.
\label{eqAL}
\end{equation}
It is important to mention that  because $\alpha$ is the local
characteristic of the inter-edge state transport it should be
independent from the contact combination. In other words, a single
value of $\alpha$ obtained from different $I-V$ curves is a test
of the consistency of our description.

The linear behavior of  experimental $I-V$ curves after the
threshold means a constant slope in Eq.~(\ref{eqAL}) and,
therefore, a constant $\alpha$. In Figs.~\ref{IV42},\ref{IV32} the
linear regions of experimental $I-V$ curves are fitted by dashed
lines. These lines are calculated from formulas like
Eq.~(\ref{eqAL}) with constant single $\alpha$ for every filling
factor combination. The used values of $\alpha$ are $0.55$ for
$\nu=4,g=2$ factors and $0.34$ for $\nu=3,g=2$.  It can be seen
from the figures, that dashed lines fit the experimental curves
quite well, even in view of small non-linearity of the
experimental curves. The same values of $\alpha$ were obtained
from similar linear fits for other samples with different gate-gap
widths. We should conclude, that $\alpha$ depends only on the
filling factor combination and therefore describe  fundamental
properties of the inter-edge-state transport.

The fact that the experimental traces are not exactly linear, see
Figs.~\ref{IV42},\ref{IV32}, indicates that there is a slow
dependence of $\alpha$ on the voltage imbalance between edge
states. Using formulas like Eq.~(\ref{eqAL}) it is possible to
extract this dependence of $alpha$ directly from the experimental
traces. In the insets to Fig.~\ref{But} the dependence of $\alpha$
is depicted as function of the voltage imbalance $V$ between edge
states  for two different filling factor combinations. Just from
the definition $\alpha$ is zero before the threshold, it jumps to
the values that close but slightly higher than ones for full
equilibration between all involved edge states ($\alpha_{eq}=1/2$
for $\nu=4,g=2$ filling factors and $\alpha_{eq}=1/3$ for
$\nu=3,g=2$)  and then  slowly diminishing while increasing the
voltage imbalance $V$ between edge states. For a single filling
factor combination $\alpha(V)$ traces obtained from different
contact configurations deviate within 2\% which is of the order of
our experimental accuracy, which also indicates the universal
character of the $\alpha$ parameter.

Let us discuss the obtained dependence of $\alpha$  on voltage
imbalance between edge states, see insets to Fig.\ref{But}. It is
important to mention that $\alpha$ by definition describes the
{\em resulting} mixing of the electrochemical potentials, while
charge transfer is taking place on the whole length of the
gate-gap width. This charge transfer is changing the
electrochemical potentials of the edge states. It means that while
at one (injection) corner of the gate-gap the energy shift between
edge states equals to the depicted in the figures voltage
imbalance $V$ between them, the edge potential profile between
edge states is flattening while moving away from the injection
corner.  At some point  the edge profile becomes to be flat. If
this point is really within  the gate-gap, full equilibration
between edge states is established,  and it should be no further
charge transfer on the rest of the gate-gap width. The resulting
value of $\alpha$ in this case can be expected to be exactly equal
to the equilibrium one. The experimental fact that the values of
$\alpha$ are higher than the equilibrium ones indicates, that
charge transfer {\em in the same direction} is still taking place
even after the equilibration point. In this case the slow
dependence of $\alpha$ on the voltage imbalance $V$ becomes clear:
at higher imbalance $V$ a higher amount of electrons should be
transferred between edge state to flatten the potential, thus the
point of equilibration moves to the opposite to the injection
corner of the gate-gap and "length of overflowing" (on which an
additional charge is transferred) becomes shorter. After leaving
the gate-gap, equilibration is not established at all, so $\alpha$
becomes smaller than the equilibrium value. This behavior can be
clearly seen in insets to Fig.~\ref{But}. The origin of the
"overflowing" behavior is still unclear and needs in further
theoretical investigations. One qualitative explanation can be
proposed here: the value of the threshold voltage $V_{th}$ is
determined by the cyclotron splitting, but not exactly it is, see
Ref.~\onlinecite{alida}. At least the energy level broadening has
an influence on the value of $V_{th}$, and maybe any other
factors. In this case we can suppose a small variation of $V_{th}$
along the gate-gap, which leads to the additional charge transfer.

It is worth to mention that for the filling factor combination
$\nu=3,g=2$ experimental values of $\alpha$ varies around the
value of $1/3$. This is the equilibrated value at which all three
edge states are involved into the charge transfer. It means that
electrons from inner edge state, having spin in the field
direction, "up", are moving both in the neighbor edge state with
spin "down" and in the outer edge state with spin "up". These
processes should go together: without high voltage imbalance,
equilibration between spin-split edge states goes on a millimeter
distance~\cite{muller}, so to have the full equilibration between
all three edge states on few microns as well process with
spin-flip should be present as one without it. (For $\nu=4,g=2$
filling factors, where the transport goes between two pairs of
equilibrated spin-split edge states, spin-flip is not needed.) At
voltages above $V_{th}$ but below $2V_{th}$ electrons are moving
by vertical relaxation through the cyclotron gap and a diffusion
in space afterwards. In the relaxation process the energy is
changing by emitting a photon (in spin-flip transfer) or a phonon
(without spin-flip). As the voltage imbalance exceeds $2V_{th}$,
the energy levels are bent enough to allow {\em horizontal}
transitions between edge states~\cite{DNP}. In these transitions
electron spin is flipping due to flopping of nuclear spin, in so
called flip-flop processes, which leads to the formation of a
nuclear polarized region in the gate-gap. This process is well
known in the literature~\cite{DNP,dixon,komiyama} as a dynamic
nuclear polarization. Once appeared, a region of dynamically
polarized nuclei influences  the electron energies through the
effective Overhauser field. Overhauser field is effectively
compensating the external field for the Zeeman splitting, and can
be in GaAs as high as 5~T, see Ref.~\onlinecite{safarov}. Thus, it
can significantly change the space distance between spin-split
edge states and therefore increase a distance for the charge
transfer in the gate-gap (which is determined by the difference
between cyclotron and spin splittings). This give rise to increase
of the resistance, once makes harder the  charge transfer. In the
experiment, it is at this voltage $V=2V_{th}$ experimental $I-V$
traces change their slopes for $\nu=3,g=2$ filling factors, see
the inset to Fig.\ref{IV32}. Also a hysteresis on the $I-V$ curves
for $\nu=3,g=2$ above the voltage $2V_{th}$ is present (not shown
in the figure), which is a key feature of the dynamic nuclear
polarization~\cite{DNP,dixon,komiyama}.

We used a quasi-Corbino sample geometry with independent contacts
to different edge states in the quantum Hall effect regime to
investigate a charge transfer between cyclotron-split edge states
at high imbalance.  We found  that charge transfer  between
cyclotron-split edge states at high imbalance can be described by
a single parameter, which is the transferred portion of the
available for transfer part of the electrochemical potential
imbalance between edge states. From the experiment we obtained
this parameter in it's dependence on the voltage imbalance between
edge states and proposed a qualitative explanation.

We wish to thank Dr. A.A.~Shashkin for help during the experiments
and Prof. A.~Lorke for valuable discussions. We gratefully
acknowledge financial support by the Deutsche
Forschungsgemeinschaft, SPP "Quantum Hall Systems", under grant LO
705/1-2.  The part of the work performed in Russia was supported
by RFBR,  the programs "Nanostructures" and "Mesoscopics" from the
Russian Ministry of Sciences. V.T.D. acknowledges support by A.
von Humboldt foundation. E.V.D. acknowledges support by Russian
Science Support Foundation.

\end{document}